\documentclass[
preprint,
11pt,
nofootinbib,
amsmath,
amssymb,
aps,
prl
]{revtex4-1}

\usepackage{bm}
\usepackage{graphicx}

\newcommand{\paperTitle}{Planar Fourier Optics for Slab Waveguides, Surface Plasmon Polaritons and 2D Materials}
\newcommand{\me}{Benjamin Wetherfield}
\newcommand{\myemail}{bsw28@cam.ac.uk}
\newcommand{\tim}{Timothy D. Wilkinson}
\newcommand{\EE}{Electrical Engineering Division, Engineering Department, University of Cambridge, UK}
\newcommand{\ABSTRACT}{
Recent experimental work has demonstrated the potential to combine the merits of diffractive and on-chip photonic information processing devices in a single chip by making use of planar (or slab) waveguides. Researchers have adapted key results of 3D Fourier optics to 2D, by analogy, but rigorous derivations in planar contexts have been lacking. Here, such arguments are developed to show that diffraction formulas familiar from 3D can be adapted to 2D under certain mild conditions on the operating speeds of the devices in question. Equivalents to the Rayleigh-Sommerfeld diffraction (RS) formulas in 2D are provided and a Radiation Condition of validity proved. The equivalence of the first 2D RS formula with an angular spectrum formulation is demonstrated. Finally Fresnel approximations are derived starting from the RS formulation and that of the angular spectrum. In addition to serving those working with slab waveguides, this letter provides analytical tools to researchers in any field where 2D diffraction is encountered, including the study of surface plasmon polaritons, surface waves, 3D diffraction with line-sources or corresponding symmetries, and the optical, acoustic and crystallographic properties of 2D materials.
}

\begin{document}

\title{\paperTitle}

% revtex
\author{\me}
\email{\myemail}
\author{\tim}
\affiliation{\EE}

% \author[1,*]{\me}
% \author[1]{\tim}
% \affil[1]{\EE}

% \affil[*]{\myemail}

% \author{\me\authormark{1, *}}
% \author{\tim\authormark{1}}
% \address{\authormark{1}\EE}
% \email{\authormark{*}\myemail}
% \date{}

% revtex
\begin{abstract}
\ABSTRACT
\end{abstract}

% osalengthcheck
% \begin{abstract}
% \ShortAbstract
% \end{abstract}

% \begin{abstract}
% \ABSTRACT
% \end{abstract}
% \setboolean{displaycopyright}{false}

% optica
% \begin{abstract*}
% \ABSTRACT
% \end{abstract*}

\maketitle

\section{Introduction}

Fourier optics is a theoretical and computational framework for understanding and simulating the propagation of light through free-space and dielectric optical elements, such as lenses, zone plates, gratings and apertures \cite{goodmanIntroductionFourierOptics2005}. Its uses and applications are many and varied, underpinning the analysis of classical imaging systems -- from microscopes to telescopes \cite{goodmanIntroductionFourierOptics2005}, the realization of super-resolution imaging via Fourier Ptychography \cite{zhengConceptImplementationsApplications2021}, the implementation of digital holography for phase-sensitive imaging and retrieval \cite{javidiRoadmapDigitalHolography2021}, the computationally tractable production of computer-generated holograms (CGH) for 3D display applications \cite{blinderStateoftheartComputerGenerated2022} and the design of optical computing devices for low energy, high-throughput and signal-integrated inferential computer vision systems \cite{xuMultichannelOpticalComputing2022}. In addition to providing a satisfactory model of electromagnetic wave propagation in many applications, Fourier optics provides highly computationally efficient formulations of diffraction by leveraging Fast Fourier Transforms (FFTs), making it ideal for training large physics-aware deep learning models \cite{shiRealtimePhotorealistic3D2021}, incorporating into intensive gradient descent procedures \cite{pengNeuralHolographyCameraintheloop2020} or performing real-time optimizations on conventional computer hardware \cite{christopherImprovingPerformanceSinglepass2020}. In other cases, the Fourier optical framework demonstrates the potential to extract useful computational operations from light propagation itself, since in specific optical layouts (such as that produced by a single lens), the action of propagation through the optical system results, to a reasonable accuracy, in a Fourier
% (or related \cite{ozaktasFractionalFourierOptics1995, wetherfieldFastDSTBasedGerchbergSaxton2021a})
relationship between input and output planes \cite{macfadenOpticalFourierTransform2017}. Alternatively, the ready availability of an optical Fourier transform can be exploited to perform Fourier domain filtering using, for example, spatial light modulators \cite{changHybridOpticalelectronicConvolutional2018} or metasurfaces \cite{silvaPerformingMathematicalOperations2014}, for the hardware acceleration of computationally expensive convolutions.

In recent work in the field of photonic computing, some of the familiar models from Fourier optics in 3D free-space have been adapted to the setting of slab (or planar) waveguides \cite{zareiIntegratedPhotonicNeural2020,ongPhotonicConvolutionalNeural2020,fuOnchipPhotonicDiffractive2021,yanAllopticalGraphRepresentation2022a,zhuSpaceefficientOpticalComputing2022,fuPhotonicMachineLearning2023,levyInhomogenousDielectricMetamaterials2007,ahmedIntegratedPhotonicFFT2020,wangDesignCompactMetaCrystal2022}, and surface plasmon polaritons \cite{kouOnchipPhotonicFourier2016}, providing diffractive processing resources in chip-sized form factors. In \cite{yanAllopticalGraphRepresentation2022a}, an angular spectrum model is used to optimize the settings of a series of 1D metasurfaces inserted between diffractive regions, implementing the weights of consecutive layers of a neural network, by analogy with work in 3D free-space \cite{linAllopticalMachineLearning2018}. In \cite{fuOnchipPhotonicDiffractive2021,fuPhotonicMachineLearning2023}, a Fourier domain Fresnel approximation is used to similar effect. By contrast, scalar diffraction can be described directly in terms of Green's functions, loosely representing the propagation of point sources or Huygens' wavelets. This style of analysis can be used to show that a star coupler architecture functions as a Fourier transforming device \cite{dragoneEfficientStarCouplers1989,zhuSpaceefficientOpticalComputing2022}, a tool that has been used to implement the convolution steps of neural networks \cite{ongPhotonicConvolutionalNeural2020,zhuSpaceefficientOpticalComputing2022}, miniaturising free-space designs that rely on lens optics \cite{changHybridOpticalelectronicConvolutional2018}.

While the down-conversion of 3D formulas into 2D equivalents has had some practical success so far in the literature, there are certain gaps in the theoretical treatments presently available. We address these gaps in this letter, providing theoretical parity with the key results of free-space optics.
In all the derivations, care is taken around the manipulation of Green's functions. In 3D, we are familiar with the field of a point-source resembling a complex exponential. Not so in 2D -- we begin with a Hankel function to represent the expanding point-source in planar contexts, and it is only after asymptotic approximations are applied that we recover a complex exponential to high accuracy for sufficiently large argument \cite{gburMathematicalMethodsOptical2011}. The following results are obtained.
First, we present equivalents to the Rayleigh-Sommerfeld (RS) diffraction formulas, from which a Fresnel approximation is derived, along with an appropriate Radiation Condition.\footnote{For corresponding results in 3D, see \protect\cite{goodmanIntroductionFourierOptics2005,gburMathematicalMethodsOptical2011}.} Second, we present a 2D angular spectrum formulation, show that it is equivalent to a direct formulation of diffraction, one of the RS formulations,\footnote{For corresponding results in 3D, see \protect\cite{shermanApplicationConvolutionTheorem1967}.} and show that it yields a Fourier domain Fresnel approximation. Third, we observe the equivalence of the Fresnel approximations in Fourier and direct forms.\footnote{For corresponding results in 3D, see \protect\cite{goodmanIntroductionFourierOptics2005}.}

\section{Results}
\label{sec:results}

For the purposes of this letter, we adopt a time-independent formulation of wave propagation, which breaks down at high operating frequencies \cite{wetherfieldFundamentalChallengesOnChip2023}, but which is consistent with the modeling assumptions of most of the existing experimental and simulation-based literature \cite{zareiIntegratedPhotonicNeural2020,ongPhotonicConvolutionalNeural2020,fuOnchipPhotonicDiffractive2021,yanAllopticalGraphRepresentation2022a,zhuSpaceefficientOpticalComputing2022,fuPhotonicMachineLearning2023,levyInhomogenousDielectricMetamaterials2007,ahmedIntegratedPhotonicFFT2020,wangDesignCompactMetaCrystal2022,kouOnchipPhotonicFourier2016}. We return to this assumption in the Discussion section. For now, we analyze the behavior of a scalar wave that is assumed to be time-separable in the sense that $u(x,z,t)= U(x, z)\exp(i\omega t)$ and $U$ satisfies a Helmholtz (or time-independent wave) equation:
\begin{equation}
  \label{eq:helmholtz_eqn}
\left (\nabla^{2}_{xz} + k^{2} \right )U(x,z) = 0
\end{equation}
where $k$ is an appropriate wavenumber. We consider $U$ in a closed region $S$ with boundary $\partial S$, subject to Neumann or Dirichlet boundary conditions.

This modelling approach is broadly applicable. For example, this approach encompasses the modeling of a slab waveguide in a steady state. For a TE mode with electric field vector $\mathbf{E}~=~(E_{x}, E_{y}, E_{z})^{T}$ we can write:
\begin{equation}
  \label{eq:helmholtz_eqn_slab_waveguide}
\left (\nabla^{2}_{xz} + \beta^{2} \right )E_{y}(x,z) = 0
\end{equation}
where $\beta$ is modal propagation constant, and $E_{y}$ has a fixed modal profile in the $y$ dimension. This one equation completely determines the electric and magnetic field: $E_{x}$ and $E_{z}$ are identically zero, while the magnetic field can be deduced from Maxwell's equations. The case of a TM mode is similar with the roles of $\mathbf{E}$ and $\mathbf{H}$ reversed \cite{okamotoFundamentalsOpticalWaveguides2006}.

\subsection{Rayleigh-Sommerfeld Solutions}
\label{sec:rayl-somm-solut-1}

Suppose $U$ satisfies a Helmholtz equation, as in equation (\ref{eq:helmholtz_eqn}). The Weber integral formula \cite{bakerMathematicalTheoryHuygens1950}, for which we provide a proof in the Supplementary Information, states
\begin{equation}
  \label{eq:integral_weber_main}
  U(\boldsymbol{r}) = \iint_{\partial S}   U(\boldsymbol{r'})\frac{\partial G(\boldsymbol{r} - \boldsymbol{r'})}{\partial n}
-\frac{\partial U(\boldsymbol{r'})}{\partial n}G(\boldsymbol{r} - \boldsymbol{r'})
  \; d\boldsymbol{r'}
\end{equation}
where $G$ is any Green's function satisfying
\begin{equation}
  \label{eq:Greens_helmholtz_2D}
  \left ( \nabla^{2}_{xz} + k^{2}\right )G(\mathbf{r} - \mathbf{r}') = \delta(\mathbf{r} - \mathbf{r}')
\end{equation}

If we can orchestrate a Green's function where $G(\mathbf{r} - \mathbf{r}')$ vanishes for $\mathbf{r}$ on the boundary $\partial S$, we are left with a 2D equivalent of the First Rayleigh-Sommerfeld (RS) Diffraction formula:
\begin{equation}
  \label{eq:29}
  U(\mathbf{r}) = \iint_{\partial S}  U(\boldsymbol{r'})\frac{\partial G(\boldsymbol{r} - \boldsymbol{r'})}{\partial n} \; d\boldsymbol{r'}
\end{equation}
Since Dirichlet boundary conditions provide data for $U$ on the boundary of the region of interest, the above formulation is suitable for homogeneous Helmholtz equations of this category.

Equivalently, the Second RS Diffraction formula is formed by causing $\partial G / \partial n$ to vanish on the boundary $\partial S$.
\begin{equation}
  U(\mathbf{r}) = -\iint_{\partial S}  \frac{\partial U(\boldsymbol{r'})}{\partial n} G(\boldsymbol{r} - \boldsymbol{r'}) \; d\boldsymbol{r'}
\end{equation}
Since Neumann boundary conditions provides data for $\partial U/ \partial n$ on the boundary of the region of interest, the above formulation is suitable for solving homogeneous Helmholtz equations of this category.

Constructing a Green's function with the appropriate vanishing conditions is simple enough if $S$ is a half-space, say $z\ge 0$. Then from $\mathbf{r} = (x,z)$, construct its reflection in $\partial S$, $\boldsymbol{\tilde{r}}=(x,-z)$. It is easy to show that $G^{-}_{2D}$ and $G^{+}_{2D}$ satisfy the conditions for the first and second RS formulas respectively if defined as follows:
\begin{align}
  G^{-}_{2D}(\boldsymbol{r} - \boldsymbol{r'}) &= \frac{1}{2}[G_{2D}(\boldsymbol{r} - \boldsymbol{r'}) - G_{2D}(\boldsymbol{\tilde{r}} - \boldsymbol{r'})]\\
  G^{+}_{2D}(\boldsymbol{r} - \boldsymbol{r'}) &= \frac{1}{2}[G_{2D}(\boldsymbol{r} - \boldsymbol{r'}) + G_{2D}(\boldsymbol{\tilde{r}} -\boldsymbol{r'})]
\end{align}

\begin{figure}[ht]
  \centering
  \includegraphics[width=0.45\textwidth]{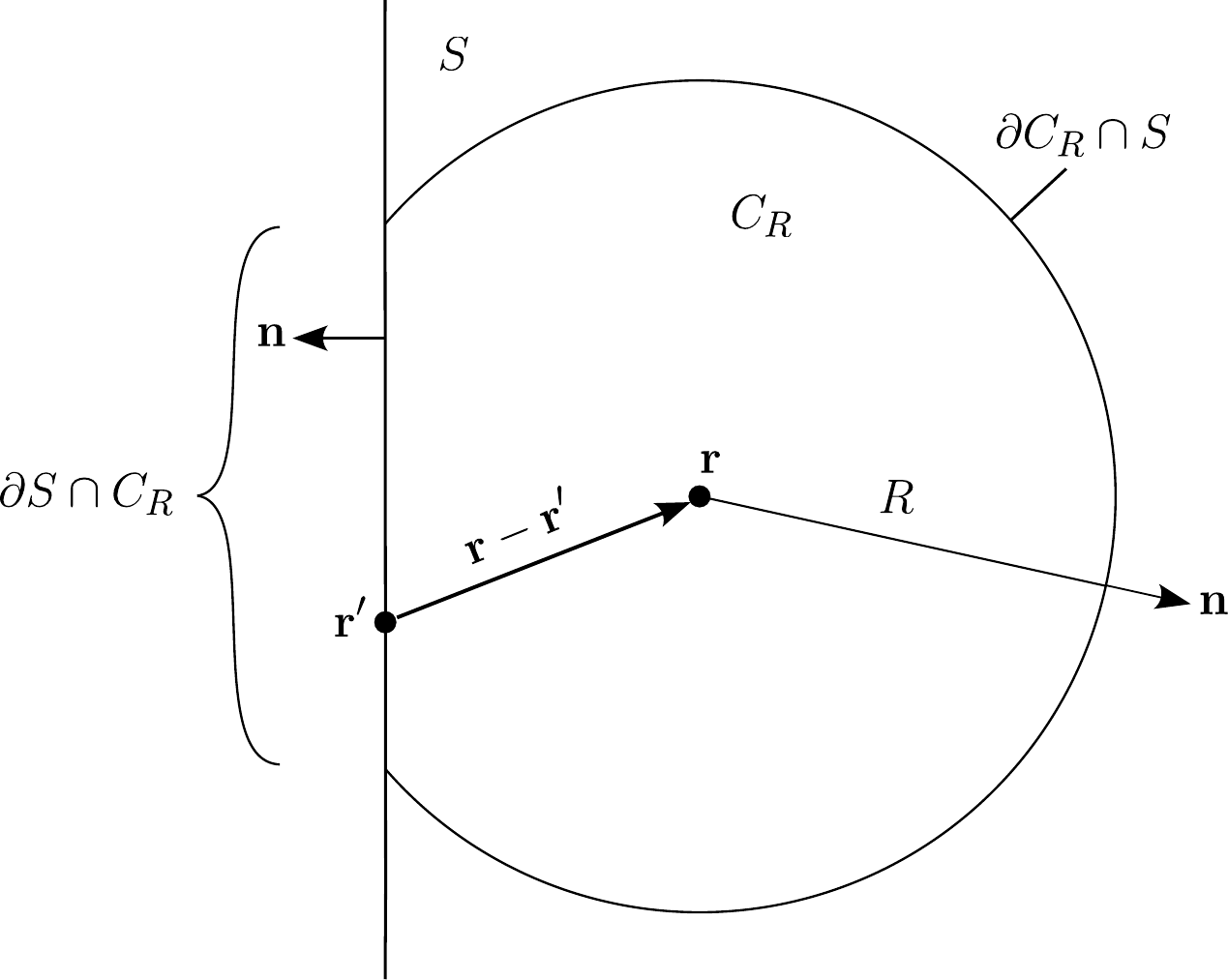}
  \caption{Construction of $S_R$ to accommodate the Radiation Condition.}
  \label{fig:radiation_condition}
\end{figure}

We have glossed over an important detail, however. The Helmholtz equation is elliptic and hence requires Dirichlet (or Neumann) conditions over a \emph{closed} boundary. We attain the half-space $z\ge 0$ as the limit as $R\to \infty$ of a parametrized  region $S_{R} = S \cap C_{R}$, where $C_{R}$ is a circle of radius $R$ centered at $\mathbf{r}$. The curved part of the boundary is given by $\partial C_{R} \cap S$, while the straight part is given by $\partial S \cap C_{R}$. One such $S_{R}$, with relevant boundary portions labeled is shown in Figure \ref{fig:radiation_condition}. By imposing the following condition (a ``Radiation Condition'') on $U$ on the boundary portion $\partial C_{R}~\cap~S$ as $R\to \infty$, we ensure that the contribution of the curved boundary to the integral tends to zero and we are free to make use of the limiting region $S$ with straight boundary at $z=0$.
\begin{equation}
  \label{eq:radiation_condition}
  \tag{Radiation Condition}
\lim_{R\to \infty} \sqrt{R}\left (\frac{\partial U}{\partial n} - ik U \right ) = 0
\end{equation}
where $U$ and $\partial U / \partial n$ are evaluated on $\partial C_{R} \cap S$, and the condition is uniform in the sense that it does not depend on angular position on the circle.

To see that the Radiation condition is sufficient to nullify the contribution of the curved boundary to the Weber integral formula, we label this contribution as $\mathcal{I}$:
\begin{align}
  \mathcal{I} &= \lim_{R\to \infty}\iint_{\partial C_{R} \cap S}  U(\boldsymbol{r'})\frac{\partial G_{2D}(\boldsymbol{r} - \boldsymbol{r'})}{\partial n} \nonumber \\
  &\quad -\frac{\partial U(\boldsymbol{r'})}{\partial n}G_{2D}(\boldsymbol{r} - \boldsymbol{r'})\; d\boldsymbol{r'}
\end{align}
% \begin{equation}
%   \mathcal{I} = \lim_{R\to \infty}\iint_{\partial C_{R} \cap S}  U(\boldsymbol{r'})\frac{\partial G_{2D}(\boldsymbol{r} - \boldsymbol{r'})}{\partial n}
%   \quad -\frac{\partial U(\boldsymbol{r'})}{\partial n}G_{2D}(\boldsymbol{r} - \boldsymbol{r'})\; d\boldsymbol{r'}
% \end{equation}
Concretely, we choose $G_{2D}$ to be
\begin{equation}
  \label{eq:concrete_greens}
G_{2D}(k(\mathbf{r} - \mathbf{r}')=-\frac{i}{4}H_{0}^{(1)}(kr)
\end{equation}
where $r=\sqrt{(x-x')^{2} + (y-y')^{2}}$. Naturally, this choice of $G_{2D}$ satisfies equation (\ref{eq:Greens_helmholtz_2D}), as needed. Applying the formula expressing the derivative of $H_{0}^{(1)}$ in terms of $H_{1}^{(1)}$, and representing the point $\mathbf{r}'$ according to its angular position on the circle, we obtain
\begin{align} \mathcal{I} &= \lim_{R\to \infty}\frac{i}{4}\bigg \{\int_{\partial C_{R} \cap S}   k U(\theta)H_{1}^{(1)}(k R) \nonumber \\
  &\quad + \frac{\partial U(\theta)}{\partial n}H_{0}^{(1)}(k R)\; d\theta \bigg \}
\end{align}
% \begin{equation} \mathcal{I}= \lim_{R\to \infty}\frac{i}{4}\bigg \{\int_{\partial C_{R} \cap S}   k U(\theta)H_{1}^{(1)}(k R)
%   \quad + \frac{\partial U(\theta)}{\partial n}H_{0}^{(1)}(k R)\; d\theta \bigg \}
% \end{equation}
Applying asymptotic expansions for $H_{0}^{(1)}$ and $H_{1}^{(1)}$:
\begin{align}
  | \mathcal{I} | &\le \frac{1}{4}\bigg |\lim_{R\to \infty}2\pi R \sqrt{\frac{2}{\pi k R}}\exp[i(kR -\pi / 4)] \nonumber \\
  &\quad\times \left (\frac{\partial U}{\partial n} -ik U \right )(1 + \mathcal{O}(R^{-1})) \bigg |
\end{align}
Being dominated by negative powers of $R$, the terms scaled by $\mathcal{O}(R^{-1})$ go to zero as $R \to \infty$. The remaining term scales in magnitude proportional to the expression in the Radiation Condition, and hence, the condition is sufficient to ensure the overall convergence of $|\mathcal{I}|$ to zero.

\subsection{Equivalence with Angular Spectrum Formulation}
\label{sec:equiv-with-angul-1}

Assuming $U$ satisfies the Radiation Condition, we can make use of the First RS Formula in the case where $S$ is the half-space $z \ge 0$. Let us suppose, as is often the case in applications, we wish to characterize $U$ at values of $\mathbf{r}$ in a plane with some fixed value $z$, propagating the wave forward from $\partial S$. Modifying notation to reflect the fact that $\partial S$ is constant in $z'$, we obtain
\begin{equation}
  \label{eq:straight_boundary}
  U(x,z) = \int_{-\infty}^{\infty}  U(x', z')\frac{\partial G_{2D}^{-}(x-x',z-z')}{\partial (-z)} \; d{x'}
\end{equation}
This equation expresses a convolution in $x$. Hence, we can apply the convolution theorem, provided we can compute the Fourier transforms of $U$ and $G_{2D}^{-}$ in the $x$ dimension. This will yield an angular spectrum formulation of wave propagation.

To characterize the $x$-spectrum of $G_{2D}^{-}$ we first put it in a different form through a series of manipulations. A similar argument appears in \cite{tyrasRadiationPropagationElectromagnetic1969}. We start with equation \eqref{eq:Greens_helmholtz_2D}, with $G_{2D}^{-}$ fulfilling the role of $G$. Applying a 2D Fourier transform to both sides, evaluating (using integration by parts), then applying an inverse Fourier transform to both sides yields
\begin{equation}
  \label{eq:28}
G_{2D}^{-}(\mathbf{r} - \mathbf{r}') = -(2\pi)^{-2}\iint \frac{\exp(i\boldsymbol{\alpha} \cdot (\mathbf{r}-\mathbf{r}'))}{ \alpha^{2}-k^{2} } \; d\boldsymbol{\alpha}
\end{equation}
where $\alpha_1$ and $\alpha_2$ are the frequency variables corresponding to $x$ and $z$, $\boldsymbol{\alpha}=(\alpha_1, \alpha_2)$ and $\alpha = \sqrt{\alpha_1^2 + \alpha_2^2}$.

Splitting away part of the integral by Fubini's theorem, we solve a contour integral in $\alpha_{2}$. If $z-z'>0$, we can form a contour in the upper half of the $\alpha_{2}$ complex plane, surrounding a simple pole at $+(k^{2} - \alpha_{1}^{2})^{1/2}$. If $z-z'<0$, we can form a contour in the lower half of the $\alpha_{2}$ complex plane, surrounding a simple pole at $-(k^{2} - \alpha_{1}^{2})^{1/2}$. Applying the residue theorem, we are left with an angular spectrum representation of $G_{2D}^{-}(\mathbf{r} - \mathbf{r}')$:
\begin{equation}
 -\frac{i}{4\pi} \int_{-\infty}^{\infty} \frac{\exp(i[\alpha_{1}(x-x')+(k^{2}-\alpha_{1}^{2})^{1/2}|z-z'|] )}{(k^{2} - \alpha_{1}^{2})^{1/2}} \; d\alpha_{1}
\end{equation}
We can now differentiate with respect to $z$ under the integral sign to obtain an angular spectrum representation of $\frac{\partial }{\partial z}G_{2D}^{-}(\mathbf{r} - \mathbf{r}')$:\footnote{It was assumed that $z>z'$, which allows us to remove the absolute value sign for the purposes of differentiation.}
\begin{equation}
  \frac{1}{4\pi} \int_{-\infty}^{\infty} {\exp(i[\alpha_{1}(x-x')+(k^{2}-\alpha_{1}^{2})^{1/2}|z-z'|] )} \; d\alpha_{1}
\end{equation}

And, by the Fourier inversion theorem,
\begin{equation}
  \label{eq:angular_spectrum}
\mathcal{F}_{x}\left \{\frac{\partial }{\partial z}G_{2D}^{-}\right \}(\alpha, z-z')=\frac{1}{2}\exp(i(k^{2}-\alpha^{2})^{1/2}|z-z'|)
\end{equation}

This is a familiar angular spectrum formulation, as used in \cite{yanAllopticalGraphRepresentation2022a}.
It is a useful computational formulation since it can be implemented in terms of FFTs.
 By use of the convolution theorem,
\begin{equation}
  \label{eq:34}
  U(x,z)=\mathcal{F}_{\alpha}^{-1}\left \{\mathcal{F}_{x}\{U\}(\alpha, z') \cdot \mathcal{F}_{x}\left \{\frac{\partial}{\partial z}G_{2D}^{-}\right \}(\alpha, z-z')\right \}
\end{equation}
where, in practice, the multiplication $\cdot$ can be vectorized.

\subsection{Fresnel Approximation}
\label{sec:fresn-appr}

To form an approximation to the integral in equation (\ref{eq:straight_boundary}), we first observe that the normal derivative of $G_{2D}^{-}$ matches that of $G_{2D}$ on the boundary $\partial S$:
\begin{align}
  \frac{\partial }{\partial z}G^{-}_{2D}(k r) \bigg |_{\mathbf{r}' \in \partial S}
%   &= \frac{1}{2}\left [\frac{\partial }{\partial z}G_{2D}(kr) - \frac{\partial }{\partial (-z)}G_{2D}(kr)\right ]_{\mathbf{r}' \in \partial S} \\
% &=\frac{\partial }{\partial z}G_{2D}(kr)\bigg |_{\mathbf{r}' \in \partial S} \\
  &= \frac{ik}{4} \frac{z - z'}{r}H_{1}^{(1)}(kr)
\end{align}

Applying an asymptotic approximation to equation \eqref{eq:straight_boundary} -- in physical terms, a small wavelength approximation -- we have
\begin{equation}
  U(x,z) \approx \frac{k}{4}e^{ -i\pi / 4}\int_{-\infty}^{\infty}  U(x', z') \sqrt{\frac{2}{\pi k r}}\exp(ikr)\frac{z-z'}{r}\; d{x'}
\end{equation}

If, further, we make a paraxial assumption, that $|x-x'|$ is small compared to $|z-z'|$, we arrive at a Fresnel approximation for 2D diffraction:
\begin{align}
  U(x,z) &\approx \frac{1}{2}\sqrt{\frac{k}{2\pi |z-z'|}}e^{ -i\pi / 4}e^{ i k|z-z'|} \nonumber \\
  &\times \int_{-\infty}^{\infty}  U(x', z') \exp\left (\frac{ik}{2}\frac{(x-x')^{2}}{|z-z'|} \right )\; d{x'}
  \label{eq:fresnel}
\end{align}

Once again, we have a convolution in the $x$ dimension. Separating out the propagation kernel, we obtain
\begin{align}
  \frac{\partial G_{2D}^{-}}{\partial z}(x,z-z') &\approx \frac{1}{2}\sqrt{\frac{k}{2\pi |z-z'|}}e^{ -i\pi / 4}e^{ i k|z-z'|} \nonumber \\
  &\quad \times \exp\left (\frac{ik}{2|z-z'|}x^{2} \right )
  \label{eq:fresnel_kernel}
\end{align}
% \begin{equation}
%   \label{eq:fresnel_kernel}
%   \frac{\partial G_{2D}^{-}}{\partial z}(x,z-z') \approx \frac{1}{2}\sqrt{\frac{k}{2\pi |z-z'|}}e^{ -i\pi / 4}e^{ i k|z-z'|}
%   \exp\left (\frac{ik}{2|z-z'|}x^{2} \right )
% \end{equation}

The Fresnel approximation can equally be expressed in the spatial frequency domain by way of the angular spectrum formulation in equation (\ref{eq:angular_spectrum}), whereupon a truncated binomial series representation of $(1-k^{2}/\alpha^{2})$ yields
\begin{equation}
  \label{eq:fresnel_spectrum}
\mathcal{F}_{x}\left \{\frac{\partial }{\partial z}G_{2D}^{-}\right \}(\alpha, z-z')\approx \frac{1}{2}e^{ik|z-z'|}\exp\left (-\frac{i|z-z'|}{2k}\alpha^{2}\right )
\end{equation}
and, it can be verified directly that equations~(\ref{eq:fresnel_kernel}) and (\ref{eq:fresnel_spectrum}) are related by a Fourier transform, as expected.

% The tour of Fourier optical results in 2D is complete, with Rayleigh-Sommerfeld solutions derived, equivalence with an angular spectrum demonstrated and equivalent Fresnel approximations emerging from each formulation respectively.

\section{Discussion}
\label{sec:discussion}

This letter has provided some of the missing analytical pieces in the use of time-independent modeling of diffraction in planar contexts. It closes the gap between prior works that may seem to take disparate approaches, but may now be seen to be equivalent or closely related. For example, it is seen that the the angular spectrum formulation used in layered diffractive photonic neural network modeling and optimization \cite{yanAllopticalGraphRepresentation2022a} can be thought of as a Fourier-domain solution to a Helmholtz equation (\ref{eq:helmholtz_eqn}). The Fourier domain Fresnel modeling approach adopted in other layered architecture optimizations is an approximate form of this same solution \cite{fuOnchipPhotonicDiffractive2021,fuPhotonicMachineLearning2023}; that this formulation is an approximate solution, increasingly accurate for larger distances, may account, in part, for the fact that the modeling was found to require longer diffraction distances to assure device accuracy (order $250\mu m$ with a Fresnel model \cite{fuPhotonicMachineLearning2023} versus distances as short as $20 \mu m$ in the angular spectrum regime \cite{yanAllopticalGraphRepresentation2022a}). In contrast to Fourier domain Fresnel approximations, derivations of the Fourier transforming properties of star couplers \cite{ongPhotonicConvolutionalNeural2020,zhuSpaceefficientOpticalComputing2022} and other curved wavefront designs \cite{kouOnchipPhotonicFourier2016} make use of a direct Fresnel model of diffraction, relying on the Weber integral formula, as we do.

The assumption that the wave function in question is time-separable as $u(x,z, t) = U(x,z)\exp(i\omega t)$, however, while a reasonable assumption in systems of low operating systems, fails to reflect the physical reality of 2D wave propagation, which is diffusive in nature, with wavefronts spreading as they propagate \cite{gburMathematicalMethodsOptical2011,courantMethodsMathematicalPhysics1962}. Indeed, for systems with high operating frequencies (in the range of 10's or 100's of GHz for standard slab waveguide configurations), a time-based error emerges in addition to other terms dependent on the spatial variables \cite{wetherfieldFundamentalChallengesOnChip2023}.

In future work, we wish to examine further distinctions between time-aware and time-independent analytical methods. The Radiation Condition derived in this work is formally necessary to find solutions for open regions, since the Helmholtz equation is an elliptic partial differential equation, which requires boundary conditions on a \emph{closed boundary} \cite{gburMathematicalMethodsOptical2011}. The radiation condition can be thought of, however, as accounting for the fact that the wave phenomena in question are temporal, and, due finite propagation speeds, will only have propagated over a finite expanse of space if measured at a given instant in time. If the full (time-dependent) wave equation, a hyperbolic partial differential equation, is employed, the requirement of a closed boundary is no longer required, and the artificial introduction of a Radiation Condition is no longer needed, yielding fewer conditions on the wave function $u$. This suggests that a time-aware approach is a more physically meaningful methodology, and indeed, we note that these considerations apply in the analysis of 3D propagation just as they do in planar contexts.

\section*{Acknowledgements}
The authors thank Ralf Mouthaan for fruitful discussions early in the process of this work.

This work was supported by the Richard Norman Scholarship grant for the Department of Engineering, University of Cambridge. For the purpose of open access, the authors have applied a Creative Commons Attribution (CC BY) licence to any Author Accepted Manuscript version arising.

\bibliographystyle{naturemag}
\bibliography{bibliography}

\begin{thebibliography}{10}
\expandafter\ifx\csname url\endcsname\relax
  \def\url#1{\texttt{#1}}\fi
\expandafter\ifx\csname urlprefix\endcsname\relax\def\urlprefix{URL }\fi
\providecommand{\bibinfo}[2]{#2}
\providecommand{\eprint}[2][]{\url{#2}}

\bibitem{goodmanIntroductionFourierOptics2005}
\bibinfo{author}{Goodman, J.~W.}
\newblock \emph{\bibinfo{title}{Introduction to Fourier Optics}}
  (\bibinfo{publisher}{{Roberts and Company publishers}},
  \bibinfo{year}{2005}).

\bibitem{zhengConceptImplementationsApplications2021}
\bibinfo{author}{Zheng, G.}, \bibinfo{author}{Shen, C.},
  \bibinfo{author}{Jiang, S.}, \bibinfo{author}{Song, P.} \&
  \bibinfo{author}{Yang, C.}
\newblock \bibinfo{title}{Concept, implementations and applications of
  {{Fourier}} ptychography}.
\newblock \emph{\bibinfo{journal}{Nature Reviews Physics}}
  \textbf{\bibinfo{volume}{3}}, \bibinfo{pages}{207--223}
  (\bibinfo{year}{2021}).

\bibitem{javidiRoadmapDigitalHolography2021}
\bibinfo{author}{Javidi, B.} \emph{et~al.}
\newblock \bibinfo{title}{Roadmap on digital holography [{{Invited}}]}.
\newblock \emph{\bibinfo{journal}{Optics Express}}
  \textbf{\bibinfo{volume}{29}}, \bibinfo{pages}{35078--35118}
  (\bibinfo{year}{2021}).

\bibitem{blinderStateoftheartComputerGenerated2022}
\bibinfo{author}{Blinder, D.}, \bibinfo{author}{Birnbaum, T.},
  \bibinfo{author}{Ito, T.} \& \bibinfo{author}{Shimobaba, T.}
\newblock \bibinfo{title}{The state-of-the-art in computer generated holography
  for {{3D}} display}.
\newblock \emph{\bibinfo{journal}{Light: Advanced Manufacturing}}
  \textbf{\bibinfo{volume}{3}}, \bibinfo{pages}{572--600}
  (\bibinfo{year}{2022}).

\bibitem{xuMultichannelOpticalComputing2022}
\bibinfo{author}{Xu, Z.}, \bibinfo{author}{Yuan, X.}, \bibinfo{author}{Zhou,
  T.} \& \bibinfo{author}{Fang, L.}
\newblock \bibinfo{title}{A multichannel optical computing architecture for
  advanced machine vision}.
\newblock \emph{\bibinfo{journal}{Light: Science \& Applications}}
  \textbf{\bibinfo{volume}{11}}, \bibinfo{pages}{255} (\bibinfo{year}{2022}).

\bibitem{shiRealtimePhotorealistic3D2021}
\bibinfo{author}{Shi, L.}, \bibinfo{author}{Li, B.}, \bibinfo{author}{Kim, C.},
  \bibinfo{author}{Kellnhofer, P.} \& \bibinfo{author}{Matusik, W.}
\newblock \bibinfo{title}{Towards real-time photorealistic {{3D}} holography
  with deep neural networks}.
\newblock \emph{\bibinfo{journal}{Nature}} \textbf{\bibinfo{volume}{591}},
  \bibinfo{pages}{234--239} (\bibinfo{year}{2021}).

\bibitem{pengNeuralHolographyCameraintheloop2020}
\bibinfo{author}{Peng, Y.}, \bibinfo{author}{Choi, S.},
  \bibinfo{author}{Padmanaban, N.} \& \bibinfo{author}{Wetzstein, G.}
\newblock \bibinfo{title}{Neural holography with camera-in-the-loop training}.
\newblock \emph{\bibinfo{journal}{ACM Transactions on Graphics}}
  \textbf{\bibinfo{volume}{39}}, \bibinfo{pages}{1--14} (\bibinfo{year}{2020}).

\bibitem{christopherImprovingPerformanceSinglepass2020}
\bibinfo{author}{Christopher, P.~J.}, \bibinfo{author}{Mouthaan, R.},
  \bibinfo{author}{Bheemireddy, V.} \& \bibinfo{author}{Wilkinson, T.~D.}
\newblock \bibinfo{title}{Improving performance of single-pass real-time
  holographic projection}.
\newblock \emph{\bibinfo{journal}{Optics Communications}}
  \textbf{\bibinfo{volume}{457}}, \bibinfo{pages}{124666}
  (\bibinfo{year}{2020}).

\bibitem{macfadenOpticalFourierTransform2017}
\bibinfo{author}{Macfaden, A.~J.}, \bibinfo{author}{Gordon, G. S.~D.} \&
  \bibinfo{author}{Wilkinson, T.~D.}
\newblock \bibinfo{title}{An optical {{Fourier}} transform coprocessor with
  direct phase determination}.
\newblock \emph{\bibinfo{journal}{Scientific Reports}}
  \textbf{\bibinfo{volume}{7}}, \bibinfo{pages}{13667} (\bibinfo{year}{2017}).

\bibitem{changHybridOpticalelectronicConvolutional2018}
\bibinfo{author}{Chang, J.}, \bibinfo{author}{Sitzmann, V.},
  \bibinfo{author}{Dun, X.}, \bibinfo{author}{Heidrich, W.} \&
  \bibinfo{author}{Wetzstein, G.}
\newblock \bibinfo{title}{Hybrid optical-electronic convolutional neural
  networks with optimized diffractive optics for image classification}.
\newblock \emph{\bibinfo{journal}{Scientific Reports}}
  \textbf{\bibinfo{volume}{8}}, \bibinfo{pages}{12324} (\bibinfo{year}{2018}).

\bibitem{silvaPerformingMathematicalOperations2014}
\bibinfo{author}{Silva, A.} \emph{et~al.}
\newblock \bibinfo{title}{Performing {{Mathematical Operations}} with
  {{Metamaterials}}}.
\newblock \emph{\bibinfo{journal}{Science}} \textbf{\bibinfo{volume}{343}},
  \bibinfo{pages}{160--163} (\bibinfo{year}{2014}).

\bibitem{zareiIntegratedPhotonicNeural2020}
\bibinfo{author}{Zarei, S.}, \bibinfo{author}{Marzban, M.~R.} \&
  \bibinfo{author}{Khavasi, A.}
\newblock \bibinfo{title}{Integrated photonic neural network based on silicon
  metalines}.
\newblock \emph{\bibinfo{journal}{Optics Express}}
  \textbf{\bibinfo{volume}{28}}, \bibinfo{pages}{36668--36684}
  (\bibinfo{year}{2020}).

\bibitem{ongPhotonicConvolutionalNeural2020}
\bibinfo{author}{Ong, J.~R.}, \bibinfo{author}{Ooi, C.~C.},
  \bibinfo{author}{Ang, T. Y.~L.}, \bibinfo{author}{Lim, S.~T.} \&
  \bibinfo{author}{Png, C.~E.}
\newblock \bibinfo{title}{Photonic {{Convolutional Neural Networks Using
  Integrated Diffractive Optics}}}.
\newblock \emph{\bibinfo{journal}{IEEE Journal of Selected Topics in Quantum
  Electronics}} \textbf{\bibinfo{volume}{26}}, \bibinfo{pages}{1--8}
  (\bibinfo{year}{2020}).

\bibitem{fuOnchipPhotonicDiffractive2021}
\bibinfo{author}{Fu, T.} \emph{et~al.}
\newblock \bibinfo{title}{On-chip photonic diffractive optical neural network
  based on a spatial domain electromagnetic propagation model}.
\newblock \emph{\bibinfo{journal}{Optics Express}}
  \textbf{\bibinfo{volume}{29}}, \bibinfo{pages}{31924--31940}
  (\bibinfo{year}{2021}).

\bibitem{yanAllopticalGraphRepresentation2022a}
\bibinfo{author}{Yan, T.} \emph{et~al.}
\newblock \bibinfo{title}{All-optical graph representation learning using
  integrated diffractive photonic computing units}.
\newblock \emph{\bibinfo{journal}{Science Advances}}
  \textbf{\bibinfo{volume}{8}}, \bibinfo{pages}{eabn7630}
  (\bibinfo{year}{2022}).

\bibitem{zhuSpaceefficientOpticalComputing2022}
\bibinfo{author}{Zhu, H.~H.} \emph{et~al.}
\newblock \bibinfo{title}{Space-efficient optical computing with an integrated
  chip diffractive neural network}.
\newblock \emph{\bibinfo{journal}{Nature Communications}}
  \textbf{\bibinfo{volume}{13}}, \bibinfo{pages}{1044} (\bibinfo{year}{2022}).

\bibitem{fuPhotonicMachineLearning2023}
\bibinfo{author}{Fu, T.} \emph{et~al.}
\newblock \bibinfo{title}{Photonic machine learning with on-chip diffractive
  optics}.
\newblock \emph{\bibinfo{journal}{Nature Communications}}
  \textbf{\bibinfo{volume}{14}}, \bibinfo{pages}{70} (\bibinfo{year}{2023}).

\bibitem{levyInhomogenousDielectricMetamaterials2007}
\bibinfo{author}{Levy, U.} \emph{et~al.}
\newblock \bibinfo{title}{Inhomogenous {{Dielectric Metamaterials}} with
  {{Space-Variant Polarizability}}}.
\newblock \emph{\bibinfo{journal}{Physical Review Letters}}
  \textbf{\bibinfo{volume}{98}}, \bibinfo{pages}{243901}
  (\bibinfo{year}{2007}).

\bibitem{ahmedIntegratedPhotonicFFT2020}
\bibinfo{author}{Ahmed, M.}, \bibinfo{author}{{Al-Hadeethi}, Y.},
  \bibinfo{author}{Bakry, A.}, \bibinfo{author}{Dalir, H.} \&
  \bibinfo{author}{Sorger, V.~J.}
\newblock \bibinfo{title}{Integrated photonic {{FFT}} for photonic tensor
  operations towards efficient and high-speed neural networks}.
\newblock \emph{\bibinfo{journal}{Nanophotonics}} \textbf{\bibinfo{volume}{9}},
  \bibinfo{pages}{4097--4108} (\bibinfo{year}{2020}).

\bibitem{wangDesignCompactMetaCrystal2022}
\bibinfo{author}{Wang, H.} \emph{et~al.}
\newblock \bibinfo{title}{Design of {{Compact Meta-Crystal Slab}} for {{General
  Optical Convolution}}}.
\newblock \emph{\bibinfo{journal}{ACS Photonics}} \textbf{\bibinfo{volume}{9}},
  \bibinfo{pages}{1358--1365} (\bibinfo{year}{2022}).

\bibitem{kouOnchipPhotonicFourier2016}
\bibinfo{author}{Kou, S.~S.} \emph{et~al.}
\newblock \bibinfo{title}{On-chip photonic {{Fourier}} transform with surface
  plasmon polaritons}.
\newblock \emph{\bibinfo{journal}{Light: Science \& Applications}}
  \textbf{\bibinfo{volume}{5}}, \bibinfo{pages}{e16034--e16034}
  (\bibinfo{year}{2016}).

\bibitem{linAllopticalMachineLearning2018}
\bibinfo{author}{Lin, X.} \emph{et~al.}
\newblock \bibinfo{title}{All-optical machine learning using diffractive deep
  neural networks}.
\newblock \emph{\bibinfo{journal}{Science}} \textbf{\bibinfo{volume}{361}},
  \bibinfo{pages}{1004--1008} (\bibinfo{year}{2018}).

\bibitem{dragoneEfficientStarCouplers1989}
\bibinfo{author}{Dragone, C.}
\newblock \bibinfo{title}{Efficient {{N}}*{{N}} star couplers using {{Fourier}}
  optics}.
\newblock \emph{\bibinfo{journal}{Journal of Lightwave Technology}}
  \textbf{\bibinfo{volume}{7}}, \bibinfo{pages}{479--489}
  (\bibinfo{year}{1989}).

\bibitem{gburMathematicalMethodsOptical2011}
\bibinfo{author}{Gbur, G.~J.}
\newblock \emph{\bibinfo{title}{Mathematical Methods for Optical Physics and
  Engineering}} (\bibinfo{publisher}{{Cambridge University Press}},
  \bibinfo{address}{{Cambridge}}, \bibinfo{year}{2011}).

\bibitem{shermanApplicationConvolutionTheorem1967}
\bibinfo{author}{Sherman, G.~C.}
\newblock \bibinfo{title}{Application of the {{Convolution Theorem}} to
  {{Rayleigh}}'s {{Integral Formulas}}}.
\newblock \emph{\bibinfo{journal}{Journal of the Optical Society of America}}
  \textbf{\bibinfo{volume}{57}}, \bibinfo{pages}{546} (\bibinfo{year}{1967}).

\bibitem{wetherfieldFundamentalChallengesOnChip2023}
\bibinfo{author}{Wetherfield, B.} \& \bibinfo{author}{Wilkinson, T.~D.}
\newblock \bibinfo{title}{Fundamental {{Challenges}} for {{On-Chip Diffractive
  Processing}} at {{Gigahertz Speeds}}} (\bibinfo{year}{2023}).
\newblock \eprint{arXiv:2303.08542}.

\bibitem{okamotoFundamentalsOpticalWaveguides2006}
\bibinfo{author}{Okamoto, K.}
\newblock \emph{\bibinfo{title}{Fundamentals of Optical Waveguides}}
  (\bibinfo{publisher}{{Elsevier}}, \bibinfo{address}{{Amsterdam ; Boston}},
  \bibinfo{year}{2006}), \bibinfo{edition}{2nd ed} edn.

\bibitem{bakerMathematicalTheoryHuygens1950}
\bibinfo{author}{Baker, B.~B.} \& \bibinfo{author}{Copson, E.~T.}
\newblock \emph{\bibinfo{title}{The {{Mathematical Theory}} of {{Huygens}}'
  {{Principle}}}} (\bibinfo{publisher}{{Oxford University Press}},
  \bibinfo{year}{1950}), \bibinfo{edition}{2nd edition} edn.

\bibitem{tyrasRadiationPropagationElectromagnetic1969}
\bibinfo{author}{Tyras, G.}
\newblock \emph{\bibinfo{title}{Radiation and Propagation of Electromagnetic
  Waves}}.
\newblock Electrical Science (\bibinfo{publisher}{{Academic Press}},
  \bibinfo{address}{{New York}}, \bibinfo{year}{1969}).

\bibitem{courantMethodsMathematicalPhysics1962}
\bibinfo{author}{Courant, R.} \& \bibinfo{author}{Hilbert, D.}
\newblock \emph{\bibinfo{title}{Methods of Mathematical Physics. 2: {{Partial}}
  Differential Equations}} (\bibinfo{publisher}{{Wiley}},
  \bibinfo{address}{{New York}}, \bibinfo{year}{1962}).

\end{thebibliography}


\begin{thebibliography}{0}%
\makeatletter
\providecommand \@ifxundefined [1]{%
 \@ifx{#1\undefined}
}%
\providecommand \@ifnum [1]{%
 \ifnum #1\expandafter \@firstoftwo
 \else \expandafter \@secondoftwo
 \fi
}%
\providecommand \@ifx [1]{%
 \ifx #1\expandafter \@firstoftwo
 \else \expandafter \@secondoftwo
 \fi
}%
\providecommand \natexlab [1]{#1}%
\providecommand \enquote  [1]{``#1''}%
\providecommand \bibnamefont  [1]{#1}%
\providecommand \bibfnamefont [1]{#1}%
\providecommand \citenamefont [1]{#1}%
\providecommand \href@noop [0]{\@secondoftwo}%
\providecommand \href [0]{\begingroup \@sanitize@url \@href}%
\providecommand \@href[1]{\@@startlink{#1}\@@href}%
\providecommand \@@href[1]{\endgroup#1\@@endlink}%
\providecommand \@sanitize@url [0]{\catcode `\\12\catcode `\$12\catcode
  `\&12\catcode `\#12\catcode `\^12\catcode `\_12\catcode `\%12\relax}%
\providecommand \@@startlink[1]{}%
\providecommand \@@endlink[0]{}%
\providecommand \url  [0]{\begingroup\@sanitize@url \@url }%
\providecommand \@url [1]{\endgroup\@href {#1}{\urlprefix }}%
\providecommand \urlprefix  [0]{URL }%
\providecommand \Eprint [0]{\href }%
\providecommand \doibase [0]{https://doi.org/}%
\providecommand \selectlanguage [0]{\@gobble}%
\providecommand \bibinfo  [0]{\@secondoftwo}%
\providecommand \bibfield  [0]{\@secondoftwo}%
\providecommand \translation [1]{[#1]}%
\providecommand \BibitemOpen [0]{}%
\providecommand \bibitemStop [0]{}%
\providecommand \bibitemNoStop [0]{.\EOS\space}%
\providecommand \EOS [0]{\spacefactor3000\relax}%
\providecommand \BibitemShut  [1]{\csname bibitem#1\endcsname}%
\let\auto@bib@innerbib\@empty
%</preamble>
\end{thebibliography}%

\end{document}

\title{\paperTitle: \\ Supplementary Information}

\author{\me}
\email{\myemail}
\author{\tim}
\affiliation{\EE}

\maketitle

\makeatletter
\renewcommand \thesection{S\@arabic\c@section}
\renewcommand\thetable{S\@arabic\c@table}
\renewcommand \thefigure{S\@arabic\c@figure}
\renewcommand \theequation{S\@arabic\c@equation}
\makeatother

\section*{General Proof of the Kirchhoff-Helmholtz-Weber Integral Formula}
\label{sec:kirchh-helmh-integr}

We prove the Weber integral formula, as stated in \cite{baker}, for two dimensional scalar diffraction, in a manner that generalizes to arbitrary dimensions since no explicit derivatives need to be taken. In three dimensions, the formula is often referred to as the Kirchhoff-Helmholtz integral formula. The formula states
\begin{equation}
  \label{eq:supp_integral_weber}
  U(\boldsymbol{r}) = \iint_{\partial S}   U(\boldsymbol{r'})\frac{\partial G_{2D}(\boldsymbol{r} - \boldsymbol{r'})}{\partial n}
-\frac{\partial U(\boldsymbol{r'})}{\partial n}G_{2D}(\boldsymbol{r} - \boldsymbol{r'})
  \; d\boldsymbol{r'}
\end{equation}
where $G_{2D}$ is a Green's function satisfying:
\begin{equation}
  \label{eq:supp_Greens_helmholtz_2D}
  \left ( \nabla^{2}_{xz} + k^{2}\right )G_{2D}(\mathbf{r} - \mathbf{r}') = \delta(\mathbf{r} - \mathbf{r}')
\end{equation}

\begin{figure}[ht]
  \centering
  \includegraphics[scale=1.5]{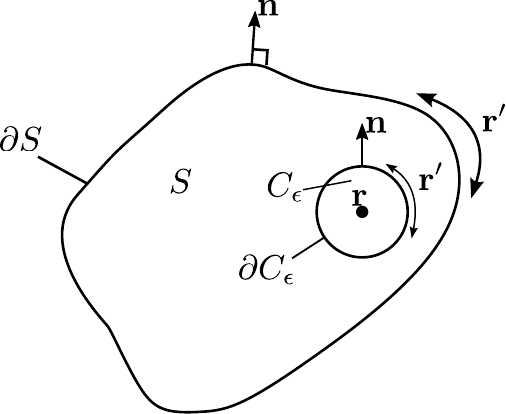}
  \caption{Diagram corresponding to the variables in the Weber integral formula and its modification to incorporate $C_{\epsilon}$.}
  \label{fig:weber_formula_diagram}
\end{figure}

To prove the formula, we introduce an auxiliary surface $C_{\epsilon}$,  a circle of radius $\epsilon$ around $\mathbf{r}$ that enables us to avoid applying the divergence theorem at a point of discontinuity. $\epsilon$ can be made arbitrarily small so that $C_{\epsilon}$ fits entirely in $S$. A diagram consisting of $S$, $C_{\epsilon}$ and the various variables in the Weber formula is presented in Figure \ref{fig:weber_formula_diagram}.

Consider the total surface $S - C_{\epsilon}$, which removes $C_{\epsilon}$ from $S$. The boundary of the reduced surface can be written $\partial S - \partial C_{\epsilon}$. In the following, we drop the subscript on $G_{2D}$, as the same argument holds in arbitrary higher dimensions just as it does in 2D (with volumes in place of surfaces). Applying the divergence theorem, then adding zero,

\begin{align}
  &\iint_{\partial S - \partial C_{\epsilon}}\left ( U(\boldsymbol{r'})\frac{\partial G(\boldsymbol{r} - \boldsymbol{r'})}{\partial n} - \frac{\partial U(\boldsymbol{r'})}{\partial n}G(\boldsymbol{r} - \boldsymbol{r'}) \right)  \; d\boldsymbol{r'}\\
  &= \iint_{S -  C_{\epsilon}} U(\boldsymbol{r}')\nabla^{2}G(\boldsymbol{r} - \boldsymbol{r'}) -G(\boldsymbol{r} - \boldsymbol{r'})\nabla^{2} U(\boldsymbol{r}')\; d\boldsymbol{r'} \\
  \label{eq:supp_6} &= \iint_{S - C_{\epsilon}}   U(\boldsymbol{r}')(\nabla^{2} + k^{2})G(\boldsymbol{r} - \boldsymbol{r'}) - G(\boldsymbol{r} - \boldsymbol{r'})(\nabla^{2} + k^{2}) U(\boldsymbol{r}')\; d\boldsymbol{r'}
\end{align}

Now, separating the surfaces $S$ and $C_{\epsilon}$,
\begin{align}
 & \iint_{S}   U(\boldsymbol{r}')(\nabla^{2} + k^{2})G(\boldsymbol{r} - \boldsymbol{r'}) - G(\boldsymbol{r} - \boldsymbol{r'})(\nabla^{2} + k^{2}) U(\boldsymbol{r}')\; d\boldsymbol{r'} \\
 &= \iint_{C_{\epsilon}}   U(\boldsymbol{r}')(\nabla^{2} + k^{2})G(\boldsymbol{r} - \boldsymbol{r'}) -G(\boldsymbol{r} - \boldsymbol{r'})(\nabla^{2} + k^{2}) U(\boldsymbol{r}')\; d\boldsymbol{r'} \\
  &= U(\boldsymbol{r})
\end{align}
by the sifting property of the delta function, and we have proved the formula.